\documentstyle[aps,twocolumn,prl,epsf]{revtex}
\begin{document}
\title{\hspace{5cm}
{\em J. Phys.: Condens. Matter} {\bf 10} {\em (1998) L265--L268}\\
\vskip 1cm
On piezophase effects in mechanically loaded atomic scale
Josephson junctions}
\author{S Sergeenkov}
\address{Bogoliubov Laboratory of
Theoretical Physics, Joint Institute for Nuclear Research\\
141980 Dubna, Moscow Region, Russia}
\address{\em (\today)}
\draft
\maketitle
\begin{abstract}
The response of an intrinsic Josephson contact to externally applied
stress is considered within the framework of the dislocation-induced
atomic scale Josephson effect. The predicted quasi-periodic (Fraunhofer-like)
stress-strain and stress-current patterns should manifest themselves for
experimentally accessible values of applied stresses in twinned crystals.
\end{abstract}
\pacs{PACS numbers: 74.50.+r, 74.62.Fj, 81.40.Jj}

It is now well-established (see, e.g.,~\cite{ref1,ref2,ref3,ref4,ref5}
and further references therein) that due to the extreme smallness of the
superconducting coherence length in high-$T_c$ superconductors (HTS),
practically any defects are capable of creating rather pronounced
weak-link structures in these materials. In particular, there are serious
arguments to consider the twinning boundary (TB) in HTS as insulating
regions of the Josephson SIS-type structure~\cite{ref6}. Given the fact
that the physical thickness of the TB is of the order of an interplane
distance (atomic scale), such a boundary should rather rapidly move via
the movement of the twinning dislocations~\cite{ref7}. And indeed, using
the method of acoustic emission, the flow of twinning dislocations
with the maximum rate of $v_0=1mm/s$ has been registered~\cite{ref8} in
$YBCO$ crystals at $T=77K$ under the
external load of $\sigma =10^7N/m^2$ (with the ultimate stress of
$\sigma _m=10^8N/m^2$).

In an attempt to describe some anomalous phenomena observed in HTS and
attributed to their weak-link structure, a rather simple model of
dislocation-induced atomic scale Josephson effect has been
suggested~\cite{ref3,ref9,ref10}. Using the above model, in this Letter we
consider the response of a single Josephson contact (created by dislocation
strain field $\epsilon _d$ acting as an insulating barrier in a SIS-type
junction) to an externally applied mechanical loading.
The resulting {\it piezophase effects}, given by stress-strain and
stress-current diagrams are found to show a typical for Josephson
junctions quasi-periodic behavior.

Recall that a conventional Josephson effect (i.e., a macroscopically
coherent tunneling of Cooper's pairs through an insulating barrier) can
be described by the following Hamiltonian
\begin{equation}
{\cal H}_J(t)=J(T)[1-\cos \phi (t)],
\end{equation}
where
$\phi (t)$ is the phase difference between two superconductors
separated by insulating layer (of thickness $l$), and
$J(T)\propto e^{-l\sqrt{U}}$ is the (temperature-dependent) Josephson
coupling energy with $U$ being a height of the insulating barrier.

According to our scenario~\cite{ref3,ref9} of dislocation-induced atomic
scale Josephson junction (JJ), the length of the SIS-type contact $L$ and
the insulator thickness $l$ supposedly correspond to the TB length and the
TB thickness, respectively.
Hence, $l$ is proportional to the number of dislocations
while $U(\epsilon _d)$ is created by defects with a strain energy
$E_d\simeq C_{44}\epsilon _d^2$ (where $C_{44}$ is the shear modulus).

As is well-known, a constant voltage $V$ applied to a JJ causes a time
evolution of the phase difference, $d\phi /dt=2eV/\hbar$, and as a result
a conventional AC Josephson current occurs through such a contact
\begin{equation}
I_s^V(t)=I_c\sin (\phi _0+\omega _Vt),
\end{equation}
where $\phi _0$ is the initial (at $t=0$) phase difference,
$\omega _V=2eV/\hbar$ the Josephson frequency, and $I_c=2eJ/\hbar$ the
critical current.

In~\cite{ref10} another possibility for AC Josephson effect was suggested
which is based on the external load ($\sigma$) induced flow of dislocations
through an unbiased ($V=0$) superconducting sample. Indeed, if we assume
that a TB (which is characterized by a non-zero dislocation-induced
strain field $\epsilon _d$) is the only source of a SIS-type Josephson
contact, then, a mechanical stress applied to such a contact will cause
a flow of dislocations through a loaded crystal, leading to the
corresponding
displacement of the insulating layer (created by these TB dislocations).
And as a result, a time-dependent phase difference
$d\phi /dt =(d\phi /d\epsilon ) \stackrel{.}{\epsilon }$
[where $\stackrel{.}{\epsilon }=d\epsilon /dt$ is the rate of plastic
deformation under an applied stress] will occur in such a {\it moving}
contact. For simplicity, a linear dependence for the induced phase
difference
$\phi (\epsilon )=A\epsilon $ (where $A\simeq 1$ is a geometrical factor)
will be assumed. Besides, to neglect any self-field effects
and to stay within a short junction approximation,
we assume (i) $L<\lambda _J$ (where $\lambda _J$
is the Josephson penetration depth), and (ii) a constant (time-independent)
rate $v_d$ of flow of dislocations through a loaded crystal.
In most cases~\cite{ref11}, $v_d\simeq v_0(\sigma /\sigma _m)$, where
$\sigma _m$ is the so-called ultimate stress. Finally, taking into
account the dependence of $\stackrel{.}{\epsilon }$ on the number of
moving dislocations (of density  $\rho $) and a mean dislocation rate $v_d$,
viz.~\cite{ref11} $\stackrel{.}{\epsilon }$=$b\rho v_d$ (here $b$ is the
absolute value of the Burgers vector), we obtain
\begin{equation}
{\cal H}_J^{\sigma}(t)=J[1-\cos (\phi _0 +\omega _{\sigma }t)],
\end{equation}
and
\begin{equation}
I_s^\sigma (t)=I_c\sin (\phi _0+\omega _{\sigma}t),
\end{equation}
for the dislocation-induced single-junction Hamiltonian and unbiased AC
Josephson current, respectively, with $\omega _{\sigma }=b\rho v_d(\sigma )$.

The response of this JJ, with an average energy
\begin{equation}
E_J(\sigma )\equiv <{\cal H}_J^{\sigma}(t)>=
\frac{1}{\tau} \int\limits_{0}^{\tau }dt {\cal H}_J^{\sigma}(t),
\end{equation}
to the externally applied tensile stress field $\sigma$ will produce
the following change $\Delta \epsilon =\epsilon -\epsilon _d$ in the
intrinsic dislocation-induced strain field $\epsilon _d$
\begin{equation}
\Delta \epsilon (\sigma )=\frac{1}{V}\frac{\partial E_J(\sigma )}{\partial
\sigma}.
\end{equation}
Here $<...>$ means a temporal averaging with a characteristic time $\tau$
(which is related to the rate of dislocation $v_0$ and its length $L$ as
$v_0\simeq L/\tau$), and $V$ is the volume occupied by TB dislocations.

To consider applied stress-induced effects only, in what follows we assume
that initially, in unloaded crystal (with $\sigma =0$),
$\epsilon (0)=\epsilon _d$ and thus $\phi _0=0$.
By resolving (3)-(6), we obtain a quasi-periodic stress-strain
relationship for a single dislocation-induced JJ
\begin{equation}
\Delta \epsilon (\sigma )=\epsilon _0f_1(\sigma /\sigma _0),
\end{equation}
with
\begin{equation}
f_1(x)=\frac{\sin x-x\cos x}{x^2}.
\end{equation}
Here $\epsilon _0=J(T)/V\sigma _0$ and $\sigma _0=\sigma _m/b\rho L$.

Let us estimate the order of magnitude of the characteristic strain
$\epsilon _0$ and stress $\sigma _0$ fields in $YBCO$ crystals.
First of all, we note that at low enough applied stress
($\sigma \ll \sigma _0$), the above model relationship
$\Delta \epsilon (\sigma )$ reduces to the linear Hooke's law,
$\Delta \epsilon =S\sigma$ with $S \equiv C_{44}^{-1}=\epsilon _0/3\sigma _0$.
Recalling~\cite{ref12} that in $YBCO$ crystals the shear modulus
$C_{44}^{exp}=8\times 10^{10}N/m^2$, we arrive at the following two equations
for the two parameters $\epsilon _0$ and $\sigma _0$, namely
(i) $3\sigma _0=C_{44}^{exp}\epsilon _0$, and (ii) $\sigma _0\epsilon _0=
J(T)/V$.
Taking $b=1.2nm$, $V\simeq b^3$ and $J(T=77K)=0.5meV$ for
the magnitude of the Burgers vector, the volume occupied
by TB dislocations, and the Josephson energy in $YBCO$
crystals~\cite{ref12}, from the above two equations we obtain
$\sigma _0\simeq 10^7N/m^2$ and $\epsilon _0\simeq 10^{-3}$.
It is worthwhile to mention that these numbers remarkably correlate with
the ones usually seen in mechanically loaded type-II hard
superconductors~\cite{ref13}.
Figure 1 shows the sress-strain diagram of an atomic scale JJ beyond the
linear approximation when rather strong plasticity effects come into action.

In the same manner, from (4) we can find the change of the Josephson
supercurrent under external loading. The resulting stress-current
relationship for a single JJ reads
\begin{equation}
\Delta I_s(\sigma )\equiv \left |\frac{1}{\tau} \int\limits_{0}^{\tau }dt
I_s^{\sigma}(t) \right |=I_cf_2(\sigma /\sigma _0),
\end{equation}
with
\begin{equation}
f_2(x)=\left |\frac{1-\cos x}{x} \right |.
\end{equation}

The behavior of the Josephson supercurrent under applied stress, given by
(9), is depicted in figure 2. The elastic Hooke's region (valid for
$\sigma \ll \sigma _0$) is characterized by a linear
dependence $\Delta I_s(\sigma )\simeq K\sigma$ (with $K=I_c/2\sigma _0$)
which can be seen in intrinsically defected samples under a relatively
small loading~\cite{ref14}. By applying a relatively small elastic stress
along the $ab$ plane of sputtered $YBCO$ thin films, Belenky
{\it et al}~\cite{ref14}
detected the following change of the critical current density
$\Delta J_c(\sigma )/J_c(0)=3\times 10^{-3}$ in their samples under
compressive load of $\sigma =10^5N/m^2$ at $T=77K$. Assuming that this change
is controlled by intrinsic inhomogeinities, we can compare it with the model
prediction which yields (see (9))
$\Delta I_s(\sigma )/I_c\simeq 5\times 10^{-3}$ for the same set of
parameters (recall that in our model $\sigma _0=10^7N/m^2$).
However, a much larger compression (close to the ultimate stress
$\sigma _m$, see above) is needed to check the validity of the both predicted
Fraunhofer-like patterns shown in figures 1 and 2.

In summary,
using a model of atomic scale Josephson contact, the change of the
defect-mediated strain field and Josephson supercurrent under external
loading was considered.
A possibility of experimental verification of the resulting stress-strain
and stress-current diagrams in defected superconductors was discussed.

\begin{figure}
\epsfxsize=8.5cm
\centerline{\epsffile{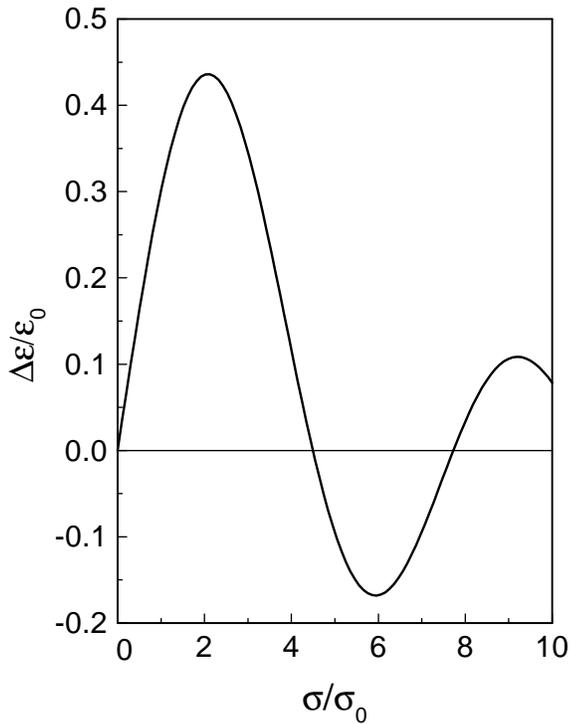} }
\caption{Stress-strain diagram of a single Josephson junction, according to
(7).}
\end{figure}

\begin{figure}
\epsfxsize=8.5cm
\centerline{\epsffile{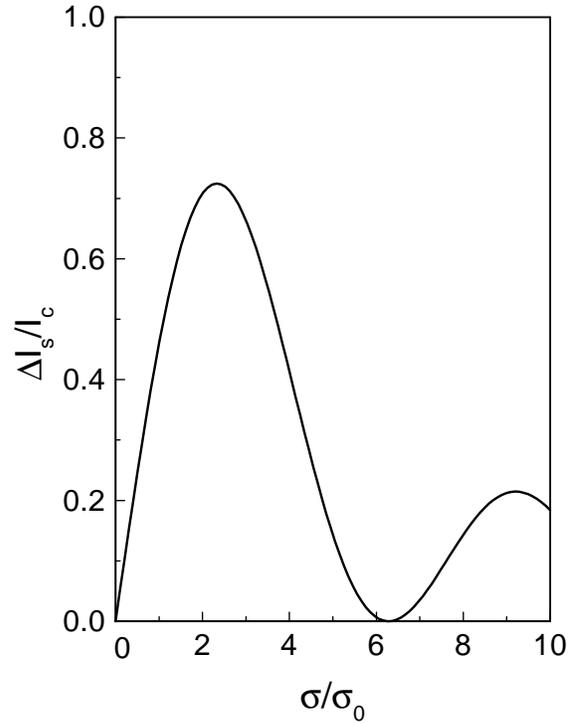} }
\caption{Stress-current diagram of a single Josephson junction, according to
(9).}
\end{figure}

\end{document}